\documentclass[twoside,english]{elsarticle}
\usepackage{ae,aecompl}
\usepackage[T1]{fontenc}
\usepackage[latin9]{inputenc}
\usepackage{geometry}
\usepackage{float}
\usepackage{amsthm}
\usepackage{amstext}
\usepackage{graphicx}

\makeatletter

\providecommand{\tabularnewline}{\\}

\theoremstyle{plain}
\newtheorem{thm}{\protect\theoremname}
\theoremstyle{definition}
\newtheorem{defn}[thm]{\protect\definitionname}

\makeatletter
\def\ps@pprintTitle{%
  \let\@oddhead\@empty
  \let\@evenhead\@empty
  \def\@oddfoot{}
  \let\@evenfoot\@oddfoot
}
\makeatother

\usepackage{fancyhdr}
\usepackage{lineno}

\usepackage{pdflscape}
\usepackage[nolist,nohyperlinks]{acronym}

\makeatother

\usepackage{babel}
\providecommand{\definitionname}{Definition}
\providecommand{\theoremname}{Theorem}

\begin{document}
\begin{frontmatter}

\title{{\Large{}RiskRank: Measuring interconnected risk}\tnoteref{mytitlenote}}

\tnotetext[mytitlenote]{The paper has benefited from presentation at the Finnish Economic
Association XXXVII Annual Meeting (KT-päivat) in February, 2015 in
Helsinki and at the RiskLab/Bank of Finland/European Systemic Risk
Board (ESRB) Conference on Systemic Risk Analytics (SRA) in September,
2015 in Helsinki. The paper is complemented with a web-based application
of networks that illustrate aggregation of risk in a hierarchical
structure: http://vis.risklab.fi/\#/fuzzyAgg. The authors thank Gregor
von Schweinitz and Tuomas Peltonen for comments and discussions. Corresponding
author: Peter Sarlin, Hanken School of Economics, Helsinki, Finland.
E-mail: peter@risklab.fi.}

\author[P3,H]{ József Mezei}

\author[P2,P3]{ and Peter Sarlin}

\address[P2]{Department of Economics, Hanken School of Economics, Helsinki, Finland}

\address[P3]{RiskLab Finland at Arcada University of Applied Sciences, Helsinki,
Finland}

\address[H]{Faculty of Social Sciences, Business, and Economics, Åbo Akademi
University, Turku, Finland}
\begin{abstract}
This paper proposes RiskRank as a joint measure of cyclical and cross-sectional
systemic risk. RiskRank is a general-purpose aggregation operator
that concurrently accounts for risk levels for individual entities
and their interconnectedness. The measure relies on the decomposition
of systemic risk into sub-components that are in turn assessed using
a set of risk measures and their relationships. For this purpose,
motivated by the development of the Choquet integral, we employ the
RiskRank function to aggregate risk measures, allowing for the integration
of the interrelation of different factors in the aggregation process.
The use of RiskRank is illustrated through a real-world case in a
European setting, in which we show that it performs well in out-of-sample
analysis. In the example, we provide an estimation of systemic risk
from country-level risk and cross-border linkages. \end{abstract}
\end{frontmatter}

\newpage{}

\section{Introduction}

The current financial crisis has stimulated research on systemic financial
risks. This has led to several contributions for measuring interconnectedness
and contagion risk, as well as for estimating probabilities of systemic
distress events. Yet, these two types of models have so far been built
in isolation. This paper proposes RiskRank as a measure of connected
risk by joining risk likelihoods and impacts.

The literature on systemic risk measurement has evolved along two
dimensions \citep{Borio2011}: cyclical and cross-sectional systemic
risk. These two dimensions accentuate the need for modeling not only
individual financial components, be they economies, markets or institutions,
but also interconnectedness among them and their system-wide risk
contributions. To this end, analytical tools and models provide ample
means for two types of tasks: (\emph{i}) early identification of vulnerabilities
and risks, and (\emph{ii}) early assessment of transmission channels
of and a system's resilience to shocks. While the first task is usually
tackled with early-warning indicators and models to derive a probability
of a systemic crisis (e.g., \citet{Alessi2011520}), macro stress-testing
models and contagion and spillover models provide means to assess
the resilience of the financial system to a wide variety of aggregate
shocks (e.g., \citet{Castrenetal2009}) and cross-sectional transmission
of financial instability (e.g., \citet{IMF}), respectively. RiskRank
aims at measuring both of these two dimensions concurrently.

In the vein of two strands of systemic risk measures, these can also
be viewed from the perspective of various approaches to aggregating
information. Early-warning models tend to focus on the aggregation
of multiple indicators into a meaningful measure of cyclical systemic
risk, which oftentimes takes the form of distress probabilities (e.g.,
\citet{Duca2012}). Further, the literature has also provided various
approaches for aggregating multiple models in order to assure more
robust model output (e.g., \citet{holopainen2015toward}). Likewise,
a large share of the literature on cross-sectional systemic risk has
focused on network-based measures of interconnectedness and connectivity
(e.g., \citet{billio2012,Peltonenetal2015}). RiskRank provides a
centrality measure for networks, but goes beyond link-based centrality
by also accounting for materialization probabilities (or node importance).

This paper puts forward RiskRank as a measure of interconnected risk.
While focusing on systemic risk, the approach is general-purpose in
nature by applying to any type of risk that exhibits individual materialization
probabilities (i.e., risk levels of components) and impact measures
(i.e., interlinkages among components ). In line with the literature
on aggregation operators, we put forward a framework motivated by
the Choquet integral as a means to aggregate risk levels to system-wide
vulnerability by also accounting for the size of interlinkages across
the components of the system (be they economies, markets or institutions).
Hence, this can also be seen as a network-based centrality measure
that also accounts for node importance (i.e., risk levels). This provides
nothing else than a likelihood of a systemic event at all levels of
the system, ranging from re-calculated risk at the lowest levels to
aggregated risk at the highest level. In this paper, we illustrate
the use of RiskRank from country-level early-warning models to connected
individual and system-wide risk. While being targeted at systemic
financial risk, this flexible tool is easily adaptable to measuring
any connected risk.

The rest of the paper is structured as follows. Section 2 discusses
systemic risk measurement and introduces aggregation operators and
centrality measures, particularly with a view to systemic risk. In
Section 3, we motivate and describe the modification of the general
form of Choquet integral into the Risk Rank measure and discuss its
most important features and use scenarios. Section 4 presents the
application of the RiskRank to the case of European systemic risk.
Finally, we conclude in Section 5.

\section{Measuring Systemic risk: A synthesis}

To quantify systemic risk, we need a broad toolbox of models to measure
and analyze system-wide threats to financial stability. In the vein
of standard risk analysis, we disentangle the topic of systemic risk
into two tasks: probability and impact. While assigning probabilities
to events aims at ranking individual risks and vulnerabilities as
per intensity (i.e., tasks of early-warning models), assessing the
severity or impact of an event complements by modeling transmission
channels and quantifying losses given their materialization. This
accentuates the need for modeling not only the likelihood of a distress
event $p_{i}^{t}$ in time $t$ for entity $i$, be they economies,
markets or institutions, but also system-wide importance by accounting
for interconnectedness and other types of transmission channels $m_{ij}^{t}$
between each entity $i$ and all other entities $j$ at time $t$.

This section discusses the role of systemic risk analysis from the
viewpoint of the cyclical and cross-sectional dimensions. We discuss
the two strands of literature for systemic risk analysis, as well
as motivate the need for a general-purpose approach for joining the
two strands.

\subsection{Systemic risk models}

Broadly speaking, tools and models can be divided into those for early
identification and assessment of systemic risks. \citet{ECB2010}
provides a mapping of tools to the following three forms of systemic
risk: (\emph{i}) early-warning models, (\emph{ii}) contagion and spillover
models, and (\emph{iii}) macro stress-testing models.

\paragraph{Cyclical systemic risk}

The first form of systemic risk focuses on the unraveling of widespread
imbalances and is illustrated by a thorough literature on the presence
of risks, vulnerabilities and imbalances in banking systems and the
overall macro-financial environment prior to historical financial
crises. This resembles Kindleberger\textquoteright s \citep{Kindleberger1996}
and Minsky\textquoteright s \citep{Minsky1982} financial fragility
view of a boom-bust credit or asset cycle. Hence, the subsequent abrupt
unraveling of the imbalances may be endogenously or exogenously caused
by idiosyncratic or systematic shocks, and may have adverse effects
on a wide range of financial intermediaries and markets in a simultaneous
fashion. Early and later empirical literature alike have identified
common patterns in underlying vulnerabilities preceding financial
crises (see, e.g., \citet{Kaminskyetal1998b} and \citet{ReinhartRogoff2008}).

First, by focusing on the presence of vulnerabilities and imbalances
in an economy, early-warning models can be used to derive probabilities
of the occurrence of systemic financial crises in the future (e.g.,
\citet{Alessi2011520} and \citet{Duca2012}). These models use a
set of vulnerability and risk indicators to identify whether or not
an economy is in a vulnerable state. The outputs of such models mostly
take the form of a probability of a crisis within a specific time
horizon and are monitored with respect to threshold values. Hence,
this provides us a probabilities of crisis $p_{i}^{t}$ in time $t$
for entity $i$, where entities may be economies, markets or institutions,
but does not provide information about the potential impact of the
individual entities on others. Typical methods used in early-warning
models include logistic models \citet{Duca2012} and machine learning
\citet{holopainen2015toward}.

\paragraph{Cross-sectional systemic risk}

The second type of systemic risk refers to two types of models for
measuring the cross-sectional dimension. Macro stress-testing models
provide means to assess the resilience of the financial system to
a wide variety of aggregate shocks, such as economic downturns (e.g.,
\citet{Castrenetal2009} and \citet{Hirtleetal2009}). These models
allow policymakers to assess the consequences of assumed extreme,
but plausible, shocks for different entities. The key question of
macro stress-testing is to find the balance between plausibility and
severity of the stress scenarios such that they are plausible enough
to be taken seriously and severe enough to be meaningful (e.g., \citet{AlfaroDrehmann2009}
and \citet{Quagliariello2009}). Third, contagion and spillover models
can be employed to assess how resilient the financial system is to
cross-sectional transmission of financial instability (e.g., \citet{IMF}).
Hence, they attempt to answer the question: With what likelihood,
and to what extent, could the failure of one or multiple financial
intermediaries cause the failure of other intermediaries? Accordingly,
this line of work provides information on system-wide importance by
accounting for interconnectedness and other types of transmission
channels $m_{ij}$ between each entity $i$ and all other entities
$j$. Yet, this provides little information on the likelihood of individual
entities being distressed.

Another type of cross-sectional systemic risk refers to a widespread
exogenous aggregate shock that has negative systematic effects on
one or many financial intermediaries and markets at the same time.
These types of aggregate shocks have empirically been shown to co-occur
with financial instabilities (see, e.g., \citet{Gorton1988} and \citet{DemirgucDetragiache1998}).
An example of such an event is the collapse of banks during recessions
due to the vulnerability to economic downturns. The third form of
systemic risk is contagion and spillover, which usually refers to
an idiosyncratic problem, be it endogenous or exogenous, that spreads
in a sequential fashion in the cross section. The cross-sectional
transmission of financial instability has been empirically shown by
a large number studies (e.g., \citet{UpperWorms2004} and \citet{LelyveldLiedorp2006}).
For instance, episodes of financial instabilities have been shown
to relate to the failure of one financial intermediary causing the
failure of another, which initially seemed solvent, was not vulnerable
to the same risks and was not subject to the same original shock as
the former. It is worth noting that contagion refers to a situation
when the initial failure is entirely responsible for subsequent ones,
whereas the term spillover is commonly used when the causal relationship
is not found or cannot be tested (see, e.g., \citet{ECB2010}). 
A recent approach presented in \citet{Tarashev2010} makes use of
the Shapley index developed originally for problems of game theory.
In the context of games, the Shapley index measures the average contribution
of a player that he/she individually generates to the group of players
as a whole. As it is described by Tarashev et al. \citet{Tarashev2010},
this can be naturally translated in the context of systemic risk analysis
to the decomposition of various measures of system-wide risks into
the systemic importance of individual entities. Lee et al. \citet{Lee2013}

\paragraph{Joining the two dimensions of systemic risk}

For the analysis of systemic risk, this provides a standard set-up
as in any type of risk analysis: The level of risk can be calculated
as the product of the probability that individual distress occurs
$p_{i}^{t}$ (e.g., $p_{i}^{t}$ that bank $i$ fails in quarter $t$)
multiplied with the severity of that event for other entities through
their interconnectedness $m_{ij}^{t}$ (e.g., the impact of bank $i$
on other banks $j$ in quarter $t$). The literature combining these
issues is scarce. A starting point has been provided by Minoui et
al. \citet{Minoiu2015} and Rancan et al. \citet{Rancan2015} by using
interconnectedness measures as predictors of crises.  Puliga et al.
\citet{Puliga2014} finds, in the case of building the network based
on Credit Default Swaps contracts, that systemic risk level estimation,
focusing on the time period before and after 2008, only increases
if macroeconomic indicators are incorporated in the construction of
the network. Yet, this provides little information on the vulnerability
of one entity and its impact on others. In this vein, Peltonen et
al. \citet{Peltonenetal2015a} have explicitly modeled the vulnerability
of one bank as a function of the vulnerability of its neighbors through
tail-dependence networks. While being a starting point, this provides
no structured approach to accounting for both dimensions simultaneously.

\subsection{Systemic risk aggregation}

In order to estimate systemic risk, we most often rely on various
approaches for aggregating information, not the least in the case
of the above mentioned two dimensions of risk: multiple indicators
into risk levels and measures of interconnectedness into network centrality.
For a more formal view to aggregation, the value is usually obtained
so that it provides a sufficient representation of the original set
and satisfies a number of predefined requirements related to the underlying
problem. For this purpose, different functions, termed as aggregation
operators or functions, can be defined that perform the task of producing
this representative value. In the following, the definitions will
be formulated for the case of aggregating values from the most common
$[0,1]$ interval (which will also later on be the used range). An
aggregation operator on $n$ arguments is a function $f:[0,1]^{n}\rightarrow[0,1]$,
which satisfies the following properties \citet{Beliakov2007}:
\begin{itemize}
\item boundary condition: if all the aggregated values are 1's (0's), then
the value of $f$ is 1 (0);
\item monotonicity: if $(x_{1},...,x_{n})\leq(y_{1},...,y_{n})$, then $f(x_{1},...,x_{n})\leq f(y_{1},...,y_{n})$
\end{itemize}
Although these two basic properties are satisfied by a wide class
of functions, in most of the use cases additional properties are required.
For instance, while the product of numbers on the $[0,1]$ interval
is an aggregation function, it suffers from the fact that the aggregated
value is smaller than the minimum of the original values. To overcome
this issue, a subclass of aggregation functions can be used \citet{Grabisch2009}:
$f$ is an averaging function if $\min(x_{1},...,x_{n})\leq f(x_{1},...,x_{n})\leq\max(x_{1},...,x_{n})$.
The application of averaging functions in economics, specifically
in systemic risk and network analysis, is mainly restricted to the
use of a handful of specifications, such as the minimum, maximum,
weighted mean, and the most commonly used arithmetic mean. The importance
of the arithmetic mean stems from the fact that it is the value with
minimum sum of squared deviations from the original aggregated numbers,
a crucial property used in many statistical methods. In different
contexts and applications, one may require the aggregated value to
satisfy a criterion different from minimizing the sum of squared deviations.
For instance, we can look at the overall financial state of a population
by employing different aggregation techniques, such as minimum, maximum,
median and average income. This section looks at the two dimensions
of systemic risk from the view of the aggregation procedures involved,
as both components of systemic risk are most often estimated using
different aggregation procedures related to indicators and network
structures.

\paragraph{Risk indicators into probability}

In the vein of the previous section on cyclical systemic risk, early-warning
indicators are utilized in various ways to obtain an estimate of risk.
This is essentially nothing else than an aggregation procedure. Herein,
we categorize them into three classes and view them from the perspective
of aggregation operators. The first class of models, denoted the signaling
approach, monitors individual indicators and identifies risk when
an indicator value exceeds a predefined threshold. Likewise, the multivariate
signaling approach monitors concurrently a set of indicators via a
performance-weighted average of several indicators, and monitors threshold
exceedances. This involves either no aggregation at all or a simple
weighted average. The second class of models makes use of different
statistical and machine learning approaches to estimate optimal weights
for combining indicators into a final probability. Some of these approaches
rely on and result in a linear aggregation (i.e., weighted average)
of the indicator values in the modeling process, with linear discriminant
analysis as one example. However, in most cases we obtain some non-linearity
in the aggregations, such as even in the simple logistic regression.
The number of different approaches employed for aggregating indicators
into a probability is large, such as classification trees \citet{Duttagupta2011},
logistic regression \citet{Duca2012}, artificial neural networks
\citet{Sarlin2014Bio} and $k$-nearest neighbors \citet{holopainen2015toward}.
The third class of estimation models relies on the combination of
methods from the second class through ensemble learning. Typical procedures
to generate ensemble models rely on estimating a large number of individual
models that are to be aggregated into a final one (for further details
on ensembles see \citet{holopainen2015toward}). Model aggregation
could happen at two different levels: aggregating binary model output
via a majority vote (i.e., median) or aggregating probabilistic model
output through arithmetic or weighted means. To this end, one can
conclude that each and every multivariate approach for deriving early-warning
models relies on an aggregation of indicators.

\paragraph{Interlinkages into centrality}

Drawing upon the literature on interconnectedness and networks, the
literature has obviously proposed a large number of measures that
aim at revealing network properties. Considering a single node in
a network, a traditional way is to look at different centrality measures
that can help to understand the role of a specific node in the network.
Centrality measures play a key role in systemic risk analysis, as
they provide means to indicate interconnectedness or overall importance
of a node. Beyond standard measures from graph theory, the literature
also consists of more customized measures, such as DebtRank \citet{Battistonetal2012}
as way of identifying systematically important nodes in a financial
system using a feedback centrality measure. Instead of aggregating
interconnections among entities, one may also approach the problem
from the perspective of decomposing risk contributions, such as the
Shapley index approach of Tarashev et al.\citet{Tarashev2010}. Their
proposal uses the Shapley index to estimate the systemic risk contribution
of entities utilizing a characteristic function defined on all the
subsets of the system of entities. They propose to use various characteristic
measures, mainly focusing on Value-at-Risk (VaR) and Expected Shortfall
(ES), but also note that any risk measure can be used as the basis
of calculating the Shapley index.

The purpose of various centrality measures is most often to describe
the nodes of a network from two perspectives: (\emph{i}) how they
affect other (e.g., neighboring) nodes directly or indirectly, and
(\emph{ii}) how they are affected by other nodes. These two different
perspectives can be measured for example with the in and out-degree
(strength) of a node. More complex measures focus on the centrality
of the nodes from different perspectives: importance of a node for
connecting others (betweenness centrality) and how distant it is on
average from all the other nodes (closeness centrality). From a general
point of view, most centrality measures essentially serve as an aggregation
measure: for every node, we collect specific information (e.g., in-degree
or shortest path) with respect to a subset of the other nodes (e.g.
neighboring nodes or the set of all nodes), and combine the information
into a single ``average value''. This corresponds to the general
notion of aggregation of summarizing a set of numerical values into
a single number that conveys some predefined characteristics of the
original set of values. According to this, a centrality measure is
a function, that assigns a real value for every node of a network
given the adjacency matrix, $A$. An important question is what information
(which sub-matrix of the adjacency matrix) should be considered in
obtaining a required characteristic of a node. For example, in-degree
of a node utilizes only the column corresponding to the node, degree
centrality makes use of one row and one column (in and out-degrees
of the node), while we potentially need the whole matrix to calculate
the shortest path or closeness centrality. Additionally, different
measures use different functions to aggregate the individual values
into a final (centrality) measure, such as the sum of the values,
the minimum operator or a combination of these two. Finally, we can
notice that formally almost all the centrality measures used in practice
satisfy the property required from a well-defined aggregation procedure:
monotonicity. For instance, if we increase the weight of a link in
the network, the degree centrality for a node or the shortest possible
distance between two nodes can never decrease.

 Most of the measures in the literature rely only on link values
when determining centrality, without considering values associated
to the nodes themselves. The purpose in the following section is to
focus on the aggregation procedure in the context of network centrality
measures and specify an operator that can incorporate both node and
link values in calculating different node and network characteristics.

\section{RiskRank as an aggregation operator}

This section describes our approach for concurrent measurement of
interconnected risk, particularly cyclical and cross-sectional systemic
risk. The considered system is represented as a directed graph, which
is based on a hierarchical decomposition of the system into an interconnected
network of the involved actors. In our model, actors can refer to
financial institutions, financial systems, countries, individual banks,
etc. As discussed in the previous section, there exists numerous approaches
to analyze networks focusing on the importance or centrality of individual
nodes and the level of interconnectedness to measure some generic
attributes of the network. To represent and assess the two discussed
dimensions of systemic risk, in the following we describe how to utilize
aggregation functions to combine information regarding both the likelihood
of individual risk and the interconnectedness within the network to
assess systemic risk. 

To present our approach for measuring systemic risk, we use the following
notations throughout this section. The system is represented as a
network with a hierarchical structure. The highest level (level 0)
of the hierarchy consists of a single node, $S$, representing the
systemic risk. The first level consists of the main components of
the system, $S_{1,1},S_{1,2},...,S_{1,n}$ (for example countries
or different financial sectors in a country). The $n$ nodes form
a complete sub-network (there is a directed link from every node to
all the other nodes), and additionally all of them are connected to
node $S$. The second level consists of $n$ complete and pair-wisely
disjoint sub-networks, each connected only to a single node from the
first level in the same manner as the first level nodes are connected
to $S$. According to this scheme, when creating a new level of the
hierarchy, a complete sub-network is created for each node in the
previous level. Additionally, there is a numeric value associated
to every node and link in the network: the likelihood of risk in the
corresponding component of the system as the node values and a measure
of impact of the starting node on the end node as the link weight.
In the following, $S_{i}$ will denote the number of nodes on level
$i$, $S_{i}^{j}$ denotes the number of nodes in the $i+1$th level
complete sub-network corresponding to node $j$ from level $i$, i.e.
$S_{i+1}=\sum_{j=1}^{S_{i}}S_{i}^{j}$, $c_{k}$ and $l_{k,j}$ will
denote the value associated to node $k$ and the weight associated
to the link between nodes $k$ and $j$, respectively. Finally, as
we are interested in the changes taking place in the system, the network
is considered at different time-points with the same structure but
different node and link values.

\subsection{From aggregation operators to RiskRank}

As we discussed above, the main goal of our model is to provide a
measure of systemic risk, and at the same time estimate the level
of vulnerability in different components of the system; as the system
as a whole itself is represented as a node, the task is to estimate
the value of a node in a future time-point given the previous values
of the nodes and links. The basic idea is that the state of a component
of a system (a node in the network), at least in a short time period,
is determined by the previous state of the given component and the
ones which directly impact it. This problem can be reformulated as
applying some kind of aggregation procedure to summarize the values
in a node and its neighbors to predict a future value of a node with
the natural choice being an averaging function. The most straightforward
solution, which can be seen as a type of degree centrality, would
bee obtained as the sum of the incoming links in the network. Although
this measure encompasses an important piece of information (the more
a component of a system is impacted by others, the more likely it
is that problems can spread through its connections), it does not
make use of the additional information on the values associated to
the nodes. Using the weighted average for example would calculate
the impact-weighted risk measure value as a generalization of the
degree centrality; the original definition is obtained in case of
all the node values equal 1. To improve this measure, one could potentially
consider the Ordered Weighted Average (OWA) operator introduced by
Yager \citet{Yager1988}. This averaging function reorders the values
to be aggregated into decreasing order and the associated weights
are assigned to the values based on this order. OWA includes, as a
special case, the $\min,$$\max,$weighted average and many of the
most used averaging functions offering more freedom in determining
the structure of the aggregation process. 

Still, there can be many applications where the features of OWA are
not sufficient to appropriately model the underlying phenomena. To
explain the most important information overlooked with the traditional
approaches, one can consider aggregation as a decision making tool:
evaluating a decision alternative by considering several criteria
with different importance. In many problems, the considered criteria
are not independent from each other, but rather a positive/negative
performance on a criterion increases/decreases the importance of other
criteria. A typical example is the case of medical diagnosis: in order
to assess a patients status, the presence of different symptoms is
checked. While the presence of different symptomes individually doest
not necessarily imply the high risk of a serious disease, the simultaneous
appearance of the symptoms indicates high risk. This interrelated
nature of decision criteria in various applications led to the utilization
of Choquet-integral based aggregation in decision making literature
\citet{LabreucheGabrisch2003}. In the problem of estimating systemic
risk using a network approach, this issue can be relevant as a component
of the system (the decision alternative) is not only affected by a
related component (criterion) directly, but also indirectly through
the joint effect of two or several connected components (interrelated
criteria). 

Before discussing the specific use of aggregation to obtain network
centrality measures to estimate systemic risk, we introduce the notion
of discrete Choquet integral as a sufficiently general averaging operator
to represent interdependencies and non-linear behavior. As we attempt
to model the joint effect of different components beyond considering
individual ones (which is a special case of a joint effect of one
component), as the basis of the aggregation, an initial function is
needed to represent these joint effects. In the most complex case,
every possible subset of the considered components (the power set)
need to be accounted for. This representation is basedon the construct
of a monotone or fuzzy measure \citep{MurofushiSugeno1989}: 
\begin{defn}
A monotone measure $\mu$ on the finite set $N=\left\{ 1,2,...,n\right\} $
is a set function $\mu:P(N)\rightarrow\left[0,1\right]$ (where $P(N)$
is the power set of $N$) satisfying the following two conditions:\end{defn}
\begin{itemize}
\item $\mu()=0,\mu(N)=1$; 
\item $A\subseteq B$ implies that $\mu(A)\leq\mu(B)$.
\end{itemize}
One of the most general formulations utilizing a fuzzy measure representation
in the aggregation process is modeled by the discrete Choquet integral
\citep{Choquet1953,Marichal2002}. To formulate the definition, we
suppose that there are $n$ number of criteria to be considered (connected
system components in our application), $c_{1},...,c_{n},$ based on
which an evaluation is performed, which results in the corresponding
$x_{1},...,x_{n}$ importance values (the impact of the components
on the considered central component). 
\begin{defn}
A discrete Choquet integral with respect to a monotone measure $\mu$
is defined as
\end{defn}
\[
C_{\mu}(x_{1},\text{\dots},x_{n})=\sum_{i=1}^{n}\left(x_{(i)}-x_{(i-1)}\right)\mu\left(C_{(i)}\right)
\]

\hspace{-0.52cm}where $x_{(i)}$ denotes a permutation of the $x_{i}$
values such that $x_{(1)}\leq x_{(2)}\leq\ldots\leq x_{(n)}$ and
$C_{(i)}=\left\{ c_{(i)},c_{(i+1)},\ldots,c_{(n)}\right\} $. 

Although it is not straightforward to see for all the cases, but all
the previously mentioned averaging functions, specifically the OWA
and consequently the weighted average, the minimum and maximum, are
all special cases of this general definition with an appropriate choice
of fuzzy measure. As a consequence of the complexity with regard to
the need of estimating potentially $2^{n}$ coefficients in the fuzzy
measure, the general Choquet integral has stared to gain popularity
in different applications only in recent years \citep{NarukawaTorra2007}.
As it was pointed out, the crucial aspect of the Choquet-integral
that makes is attractive in many applications compared to the traditional
averaging functions is the capability of modeling interaction in the
aggregation process. In the general multi-criteria formulation of
the application of Choquet-integral, this can be translated to the
interrelation of criteria: an alternative is evaluated differently
based on criterion $A$ in the simultaneous presence of criterion
$B$ than considering only $A$ without $B.$In the proposed application,
the risk level of an entity (country, bank) will be estimated based
on its present risk level and its connection to other entities. 

To ease on the complexity of estimating the joint effect of all the
considered criteria (connected components in the system) but still
moving beyond the simple representation of assuming independence,
one can consider only the joint effect of specific subsets and assume
that the others are negligible. A straightforward approach widely
discussed in the literature is the case of 2-additive Choquet integral.
In this formulation, only the pairwise interaction between criteria
(the joint effect of two connected components) is considered additionally
to the individual effects. The Choquet integral in this case can be
formulated as (see \citet{Grabisch1997}):
\[
C_{\mu}(x_{1},\text{\dots},x_{n})=\sum_{i=1}^{n}(v\left(c_{i}\right)-\frac{1}{2}\sum I(c_{i},c_{j}))x_{i}+\sum_{I(c_{i},c_{j})>0}I(c_{i},c_{j})\min(x_{i},x_{j})+\sum_{I(c_{i},c_{j})<0}\mid I(c_{i},c_{j})\mid\max(x_{i},x_{j}).
\]

The interaction measure $I(c_{i},c_{j})$ can be defined by transforming
the measure for pairs into the $\left[-1,1\right]$, a detailed discussion
can be found for example in \citet{Marichal2000}. As it was pointed
out for example in \citet{Keeney1976}, the interaction index and
the importance of a criteria can be interpreted in the context of
game theory using traditional utility theory, and the formulas can
be derived on that foundation as a form of the Shapley index. The
formula can be derived by using the concept of the Möbius transformation
of a fuzzy measure \citet{Grabisch2000}. The Shapley-index is defined
as
\[
v(x_{i})=\sum_{K\subset X\setminus i}\frac{(n-|K|-1)!|K|!}{n!}\left(\mu(K\cup x_{i})-\mu(K)\right)
\]
The Shapley index can be interpreted as the average contribution of
the node $i$ for all the possible subsets of the points including
$i$, moreover it takes its value from the $[0,1]$ interval and $\sum_{i=1}^{n}v_{i}=1$.
In our analyzed problem this translates to the overall impact a node
has on another node, directly (through a link) or indirectly (as a
consequence of interaction with another node). For example, in the
simplest case, a node is only connected to the central/analyzed node
through a direct link, with not other path between the two nodes.
In this case, the only non-zero term in calculating the Shapley index
will be when $K=\textrm{\ensuremath{\emptyset}}$, and the formula
is simplified to $\mu(x_{i})/n$. For our purpose, the important issue
is to understand the differences between the types of interaction
represented by the terms in the formula above, and how interaction
can be translated to the case of our network representation and systemic
risk analysis. In our 2-additive model, additionally to the individual
importance of a node connected to a central node to be assessed, we
want to account for the joint effect of two nodes on the central node.
In terms of the network, it means that we consider paths with length
2 that end in the central node with positive joint effect, and every
other path or other subset of edges is considered with 0 interaction
value. This corresponds to the case of no interaction or $I(c_{i},c_{j})=0$.
Negative interaction in the sense of the Choquet integral represents
the cases of disjunctive effects, which are not present in our model:
an increase in one node value will affect the central node also indirectly
through the path including the other node. This implies that in our
aggregation process, we consider only positive interaction values
in the from of the product of the . Additionally, in our application
the minimum operator contradicts the intuitive idea of risk spreading
throughout the network along the paths; while a positive interaction
indicates conjunctive behavior, using the minimum operator, an increase
in the higher node value will not affect the joint effect of the two
nodes. For this reason, we further modify the above formula by replacing
the minimum operator by the multiplication operator.

\subsection{RiskRank}

To see the correspondence between the proposed Choquet integral and
the Shapley index approach of Tarashev et al.\citet{Tarashev2010},
we first note that the most general form of the Choquet integral requires
the fuzzy measure to be specified on all the subsets of the set of
considered entities. In this sense, the fuzzy measure can be seen
as an example of the characteristic measure referred to by Tarashev
et al.. However, in the general case the calculation of the Choquet-based
aggregation cannot be simplified into a function of the Shapley index
of the individual entities. To be able to utilize the Shapley index,
we can restrict the measurable interlinkages to pairs of individual
entities, meaning that we would not define the risk contribution of
subsets with cardinality higher than 2. While this is a simplification
compared to a full utilization of the Choquet integral and the Shapley
index approach, it allows for a natural representation in the form
of a network and ensures the (2-)additivity of the proposed measure.\footnote{As we will see in Section 4, the indirect effects (i.e. the effect
resulting from the interconnectedness of two entities) are in general
negligible compared to direct effects.} Additionally, our approach allows for the decomposition of the individual
contribution of an entity further into direct and indirect effects.
This offers a deeper understanding of the risk structure within the
system. At the same time, additionally to an interconnectedness measure,
we are also concerned with incorporating an individual risk level
of the components under the analysis.

Based on the above discussion, a function resembling the structure
of the discrete 2-additive Choquet integral will be used to aggregate
values in a network to estimate systemic risk. For risk level (node
value) $x_{i}$ and interlinkage $I(c_{i},c_{j})$ between nodes $i$
and $j$ combining the interlinkage values between node $i$ and the
target node and between node $i$ and $j$, RiskRank is defined as
\[
RR(x_{1},\text{\dots},x_{n})=\sum_{i=1}^{n}\underbrace{(v\left(c_{i}\right)-\frac{1}{2}\sum_{j}I(c_{i},c_{j}))x_{i}}_{\mbox{Direct effect of component \emph{i}}}+\sum\underbrace{I(c_{i},c_{j})\prod(x_{i},x_{j})}_{\mbox{Indirect effect of component \emph{i}}}.
\]
The components of the formula express the two ways a node affects
another one. The RiskRank function estimates risk level for a specific
node. Accordingly, the notation $C_{\mu}^{S_{t}}$should be used in
general as the RiskRank of node $S_{t}$, but in the following, we
will not use the indexing unless it is of importance for which node
the RiskRank is calculated. RiskRank is calculated and interpreted
slightly differently for the central node $S,$and any other nodes
in the network; in the following we discuss the differences between
the two cases.

One important use of the RiskRank measure in analyzing the described
hierarchical network is to assign a value to the node on the top level,
$S$, representing the level of systemic risk. In this case, as this
specific node does not have an initial value, the RiskRank formula
can be straightforwardly applied: the nodes that are either connected
to $S$ or there is a path of length 2 from the node to $S$ are considered
in the calculations. In case of nodes from the second level of the
hierarchy, as they are only connected to a single node on the first
level, there is only one interaction term, while for the nodes in
the first level, as they form a complete sub-network, there are $t-1$
interaction terms for every node where $t$ is the number of nodes
on the first level. The final value for the node $S$ provides and
estimation of the likelihood that a system-wide risk is present in
the network. 

As for the other nodes in the network, usually we have a node value
(risk level) assigned before performing the aggregation process, for
example base don market estimations. According to this, additionally
to their incoming links, we need to account for this observed node
value in the aggregation process. As a straightforward solution, this
can be done by calculating the weighted average of this node value
and the value obtained from the RiskRank function based on the connections
of the node in the network. Alternatively, to keep a uniform formalism,
we can use the RiskRank equation without any additional aggregation
by introducing a new link in the network: a self-loop for the evaluated
node. The link weight is determined based on the overall exposure
of the corresponding system component to the elements of the system;
the measure of exposure can be different in different applications.
The interaction of the evaluated node and other nodes is set as 0,
in order to prevent the additional link to contribute to indirect
effect as part of paths of length 2 starting form a neighboring node.
We can write the formula as

\[
RR_{c}(x_{1},\text{\dots},x_{n},x_{c})=\sum_{i=1}^{n+1}\underbrace{(v\left(c_{i}\right)-\frac{1}{2}\sum_{j}I(c_{i},c_{j}))x_{i}}_{\mbox{Direct effect of component \emph{i}}}+\sum_{i,j}\underbrace{I(c_{i},c_{j})\prod(x_{i},x_{j})}_{\mbox{Indirect effect of component \emph{i}}}
\]
\[
=\underbrace{v(c)x_{c}}_{\mbox{Individual effect of component \emph{c}}}+\sum_{i=1}^{n}\underbrace{(v\left(c_{i}\right)-\frac{1}{2}\sum_{j\neq i}I(c_{i},c_{j}))x_{i}}_{\mbox{Direct effect of component \emph{i }on c}}
\]

\[
+\underbrace{\sum_{i}^{n}\sum_{j\neq i}^{n}I(c_{i},c_{j})\prod(x_{i},x_{j})}_{\mbox{Indirect effect of component \emph{j via i }on c}}
\]
where $c$ is the evaluated central node and $x_{c}$ is its associated
node value. The obtained number is an estimation of the ``true amount''
of risk attributable to the node and it can be utilized as an estimation
of the risk in a future time point. A further modification of the
aggregation procedure could be to specify the weight $v(c)$ as 1
instead of the value calculated by considering the network structure
and the weights of the links ending at $c$. The main motivation for
this is that, in general, we would not like the new estimation to
be smaller as a consequences of a node being largely interconnected
to other nodes of the network. For a node that has lot of connection
to other nodes with low risk level, the original formula would overemphasize
the interconnectedness, and as a consequence the individual risk level,
even from a very high starting value, could significantly decrease.
In this case it is possible that the above formula results in estimation
greater than one. For this reason, when applying this modified weight
when estimating the level of risk associated to a node, the final
value of RiskRank should be calculated as $\min(RR_{c},1)$. 

The proposed measure can be extended to account for more complex interaction
effects. In the above discussion, paths with length not greater than
2 were considered as the potential set of connections affecting the
analyzed node. By modifying the formula and considering paths with
length greater than 2, we can account for indirect effects that may
take into consideration the speed of the risk spreading throughout
the system. By accounting for indirect effects from nodes that can
reach the central node on a path with length at most $k$, one can
estimate the level of risk by adjusting for a longer future time period.
In this respect, the aggregated systemic risk obtained in the example
evaluates a situation that will take place in a future time-point;
if we assume that the delay of spreading risk from one node to a neighboring
node is one time unit, then the example estimates the development
of the system in two time units from now. Formally, to estimate the
state of the system $k$ time units from now based on the network
representation at time point $t$, a different RiskRank measure can
be defined by considering non-zero interaction values for nodes along
a paths that have length not greater than $k$ and has the node for
which we are estimating the risk level as the endpoint of the path.
Based on each measure we can obtain an estimation of the risk values
in different nodes of the network at any point between now and $k$
time units later. As in the case of paths of length 2, an even more
important problem here is to define the interaction values to combine
the value associated to the edges on paths with different lengths.
In the original formulation of Choquet integral, this can be done
by specifying a fuzzy measure $\mu_{t}^{k}$ to be a $k$-additive
monotone measures described in \citep{Grabisch1997} and moving beyond
the complexity of the 2-additive Choquet integral. For instance, considering
the product of values as the final path value for large $k$'s in
general results in low effect values meaning that after a point we
do not gain any new information by increasing the possible length
of considered paths, while using the maximum of the individual values
on the path would increase the systemic risk value significantly after
every step. It always depends on the understanding of the underlying
domain to determine for how many steps it is still meaningful to forecast
based on the present situation. In highly fluctuating and rapidly
changing systems the recommended value should be lower than in rather
stationary systems.

\section{RiskRank: An application to Europe}

This section illustrates the use of RiskRank to aggregate risk in
a European setting. It is worth noting that RiskRank does not require
specific definitions of ``links'' and ``individual risk'', but
is rather open to any definition of the two measures in order to arrive
at a final combined aggregate. The application shown herein aims at
a final target of a European aggregate. The application provides aggregations
to higher level in the hierarchy (from individual countries to Europe),
and accounts for the interconnections in measuring risk at the same
level. To test for the added value of RiskRank relative to only measuring
individual risk, we show in the following performance comparisons
in out-of-sample tests. We first describe the out-of-sample exercises
and measures, and then move to the application.

\subsection{Evaluating model performance}

To judge the extent to which one measure outperforms another in a
realistic setup, we need sufficiently sofisticated and carefully designed
exercises and metrics. The evaluation exercises need to both measure
the quality of signals when applied a realistic setup and measure
performance with measures that mimic the problem at hand. This paper
uses recursive real-time out-of-sample tests assess performance. In
practice, this implies the use of a recursive exercise that derives
a new model at each quarter using only information available up to
that point in time. By accounting for publication lags and using information
in the manner of an increasing window, this enables in our cases testing
whether a measure would have provided means for predicting the global
financial crisis of 2007--2008, and how measures are ranked in terms
of performance for the task. 

Following the standard evaluation framework for early-warning models
in \citet{Sarlin2013b}, we aim at mimicking an ideal leading indicator
$C_{n}(h)\in\left\{ 0,1\right\} $ for observation $n$ (where $n=1,2,\ldots,N$)
and forecast horizon $h$. This implies nothing else than a binary
indicator that is one during vulnerable periods and zero otherwise.
For detecting events $C_{n}$, we need a continuous measure indicating
membership in a vulnerable state $p_{n}\in\left[0,1\right]$, which
is then turned into a binary prediction $B_{n}$ that takes the value
one if $p_{n}$ exceeds a specified threshold $\tau\in\left[0,1\right]$
and zero otherwise. The correspondence between the prediction $B_{n}$
and the ideal leading indicator $C_{n}$ can then be summarized into
a so-called contingency matrix that assigns every classification into
one of four classes: true positives (TP, correct signals times of
crisis), true negatives (TN, correct silence in tranquil times), false
positives (FP, false alarms) and false negatives (FN, missed crisis).
In terms of the elements of the contingency matrix, we can differentiate
between two different types of classification errors that a decision
maker may be concerned with: missing crises and issuing false alarms.
To formulate the concepts of usefulness and relative usefulness as
measures of classification performance in \citet{Sarlin2013b}, we
define type I errors as the share of missed crises to the frequency
of crises, i.e. $T_{1}=FN/(FN+TP)$, and type II errors as the share
of issued false alarms to the frequency of tranquil periods, i.e.
$T_{2}=FP/(TN+FP)$. Further, we need two terms: policymakers' relative
preference between type I and II errors ($\mu$) to account for the
potentially imbalanced costs of errors and the unconditional probabilities
of crises $P_{1}$and tranquil periods $P_{2}$ to account for the
potential difference in the size of the two classes. Based on these
values, we can define the loss function as:
\[
L(\mu)=\mu T_{1}P_{1}+(1-\mu)T_{2}P_{2}.
\]
Further, based on this loss function, the absolute usefulness of the
prediction model can be specified by comparing it to using the best
guess of a policymaker (always or never signaling depending on class
frequency and preferences):
\[
U_{a}(\mu)=\min(\mu P_{1},(1-\mu)P_{2})-L(\mu).
\]
Finally, we compute relative usefulness, $U_{r}$ to compare the absolute
usefulness of the model to the absolute usefulness of a model with
perfect performance ($L(\mu)=0$). Additionally, to assess predictive
performance, we also calculate standard measures from the classification
and machine learning literature, in particular the area under the
receiver operating characteristic ($ROC$) curve ($AUC$). These techniques
provide both measures tailored to the preferences of a policymaker
as well as more general-purpose measures to assess model performance.
Other performance measures to be used in assessing the model include:
(i) precision of signals $TP/(FP+TP)$, i.e. the share of correct
signals to the frequency of signals; (ii) precision of tranquil predictions
$TN/(FN+TN)$, i.e. the share of correct silence in tranquil times
to the frequency of predicting tranquil time; (iii) recall of signals
$TP/(FN+TP)$, i.e. the share of correct signals to the frequency
of crisis times; (iv) recall of tranquil predictions $TN/(FP+TN)$,
i.e. the share of correct silence in tranquil times to the frequency
of tranquil times; (v) accuracy of the model $TP+TN/(FN+FP+TN+TP)$,
i.e. the share of correct classifications.

\subsection{RiskRank for Europe and its countries}

This section measures systemic risk for individual European countries
as well as aggregates to a pan-European level. In this application,
we focus on individual risk as measured with country-level macro-financial
imbalance indicators and real linkages as measured with real cross-border
linkages. After describing the underlying data and models, we herein
both evaluate model performance and exemplify model output with and
without RiskRank. 

To be able to compute RiskRank, we need to build a network of countries
that includes both node and link values. We hence specify risk levels
(probabilities $x_{c}$) for countries, and also measures of interconnectedness
between all pairs of nodes (interlinkages $I(c_{i},c_{j})$). The
approach for computing individual risk for each economy follows \citet{holopainen2015toward}.
To derive individual risk with early-warning models, we need crisis
events and vulnerability indicators. The crisis events are based upon
the IMF database by \citet{LaeevenValencia2008}. The vulnerability
indicators used include most common measures of widespread imbalances,
such as excessive credit growth, excessive increases in stock and
house prices, GDP growth, loans to deposits and debt service ratio,
as well as more structural indicators, such as government debt, current
account deficits and inflation. We use a standard logistic regression
with 14 macro-financial indicators for countries and a forecast horizon
of 5--12 quarters prior to crisis events, as is common in the literature.
The network dimension is measured with BIS International Banking Statistics
in terms of foreign claims of banking sectors on other economies.
From the perspective of systemic banking crises, this provides ample
means to capture the spread of vulnerabilities across borders through
real linkages. Yet, these are only one means to measure both the cyclical
and cross-sectional dimensions of systemic risk and hence obviously
also come with a number of deficiencies. It is hence worth noting
that the models and data are given and the key focus is hence on relative
performance of RiskRank versus standalone early-warning models.

\begin{figure}[H]
\begin{centering}
\includegraphics[scale=0.65]{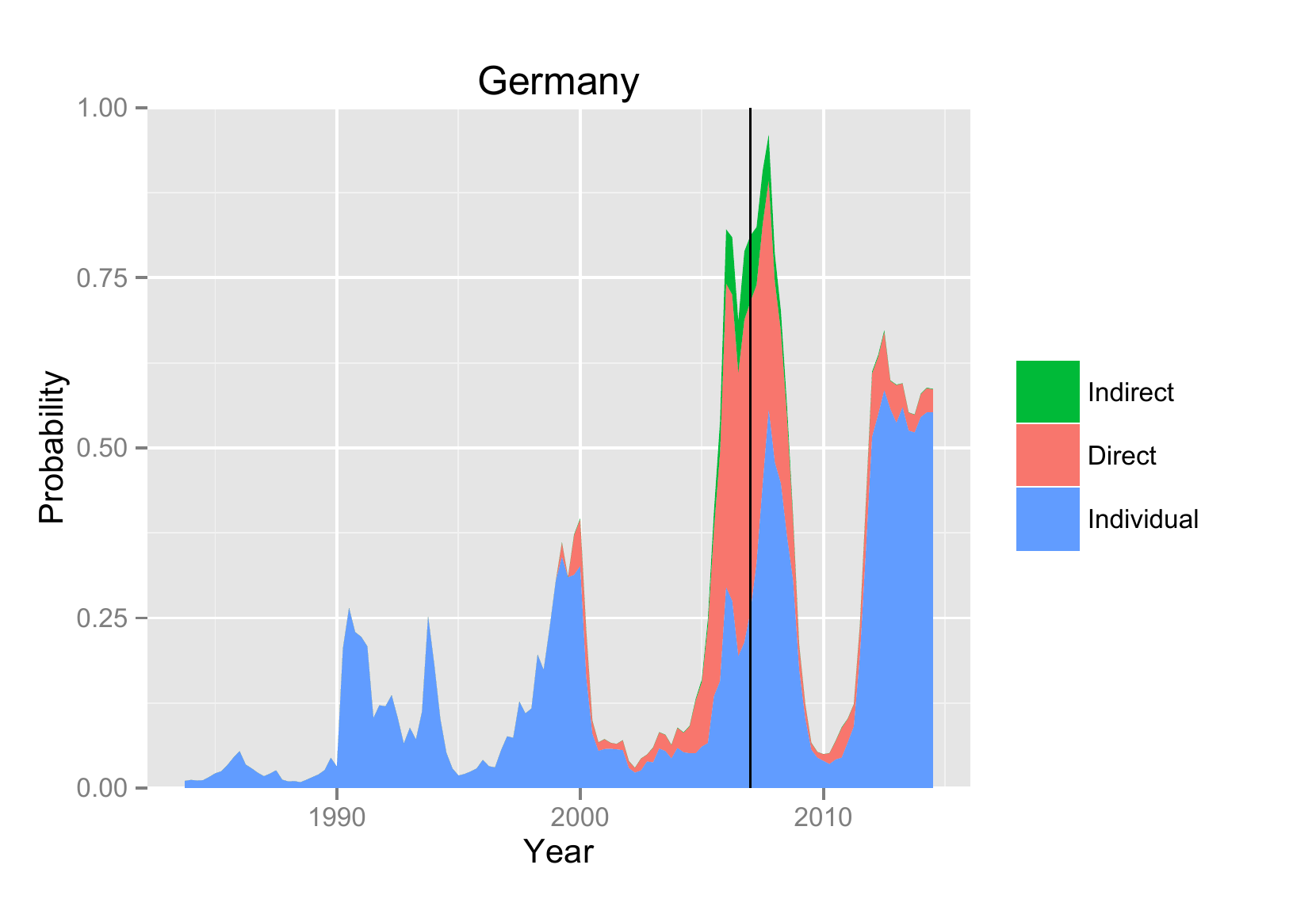}\global\long\def\newmacroname{}

\par\end{centering}

\protect\caption{\label{fig:Country_examples}Examples of individual risk and RiskRank
for Germany and Europe}
\end{figure}

To start with, we first estimate individual risk and RiskRank for
each economy separately. In the case of RiskRank, this involves aggregating
at the same level of hierarchy by also accounting for interconnectedness,
whereafter we evaluate the performance of the individual risk model
and RiskRank vis-á-vis the crisis event database. This provides the
results in Table \ref{tab:Country_Results}. As has been pointed out
in several works (e.g., \citet{Sarlin2013b,Betzetal2014}), a feature
of samples with imbalanced classes is that one needs to be more concerned
of missing crises for models to be Useful. Given preferences $\mu$,
a pairwise comparison shows that the performance of RiskRank is never
worse than the traditional individual risk model and it is better
for preferences $\mu\in\left[0.3,1\right]$. Figure \ref{fig:Country_examples}
provides an example of model output for Germany with the individual
model and RiskRank. The risk level is decomposed into individual risk,
direct and indirect effects of other countries in case of individual
countries. As we can observe from the figure, indirect effects are
most often negligible compared to the other components, and while
individual risk usually dominates the aggregated value, especially
in crisis periods the direct impact from connected countries can significantly
increase the risk level. More specifically, the timing of crisis signals
come at a much earlier stage when accounting for direct and indirect
effects. This represents the fact that domestic imbalances were fairly
modest, while Germany was highly interlinked to countries with large
vulnerabilities. As an early indication of an impending crisis, the
figure illustrates the importance to incorporate vulnerabilities descending
from ``neighboring'' countries, in addition to those building up
in the domestic economy. Likewise, one can observe that the latest
increases in risk are on the other hand related more to domestic imbalances,
which again points to the potential need for and type of domestic
policy actions.

\begin{table}[H]
\protect\caption{\label{tab:Country_Results}Signaling performance of individual models
and RiskRank for countries with forecast horizon 5--12 quarters prior
to crisis events.}

\centering{}%
\begin{tabular}{ccccc}
 & \multicolumn{2}{c}{{\small{}Individual}} & \multicolumn{2}{c}{{\small{}RiskRank}}\tabularnewline
{\small{}$\mu$} & \textbf{\small{}$U_{r}(\mu)$} & {\small{}AUC} & \textbf{\small{}$U_{r}(\mu)$} & {\small{}AUC}\tabularnewline
\hline 
\hline 
{\small{}0.0} & {\small{}0 \%} & {\small{}0.915} & {\small{}0 \%} & {\small{}0.921 }\tabularnewline
{\small{}0.1} & {\small{}-6 \%} & {\small{}0.915} & {\small{}-6 \%} & {\small{}0.921}\tabularnewline
{\small{}0.2} & {\small{}-3 \%} & {\small{}0.915} & {\small{}-3 \%} & {\small{}0.921}\tabularnewline
{\small{}0.3} & {\small{}6 \%} & {\small{}0.915} & {\small{}7 \%} & {\small{}0.921}\tabularnewline
{\small{}0.4} & {\small{}12 \%} & {\small{}0.915} & {\small{}18 \%} & {\small{}0.921}\tabularnewline
{\small{}0.5} & {\small{}15 \%} & {\small{}0.915} & {\small{}38 \%} & {\small{}0.921}\tabularnewline
{\small{}0.6} & {\small{}25 \%} & {\small{}0.915} & {\small{}39 \%} & {\small{}0.921}\tabularnewline
{\small{}0.7} & {\small{}44 \%} & {\small{}0.915} & {\small{}54 \%} & {\small{}0.921}\tabularnewline
{\small{}0.8} & {\small{}60 \%} & {\small{}0.915} & {\small{}66 \%} & {\small{}0.921}\tabularnewline
{\small{}0.9} & {\small{}73 \%} & {\small{}0.915} & {\small{}74 \%} & {\small{}0.921}\tabularnewline
{\small{}1.0} & {\small{}0 \%} & {\small{}0.915} & {\small{}0 \%} & {\small{}0.921}\tabularnewline
\end{tabular}
\end{table}

\begin{figure}[H]
\centering{}\includegraphics[scale=0.55]{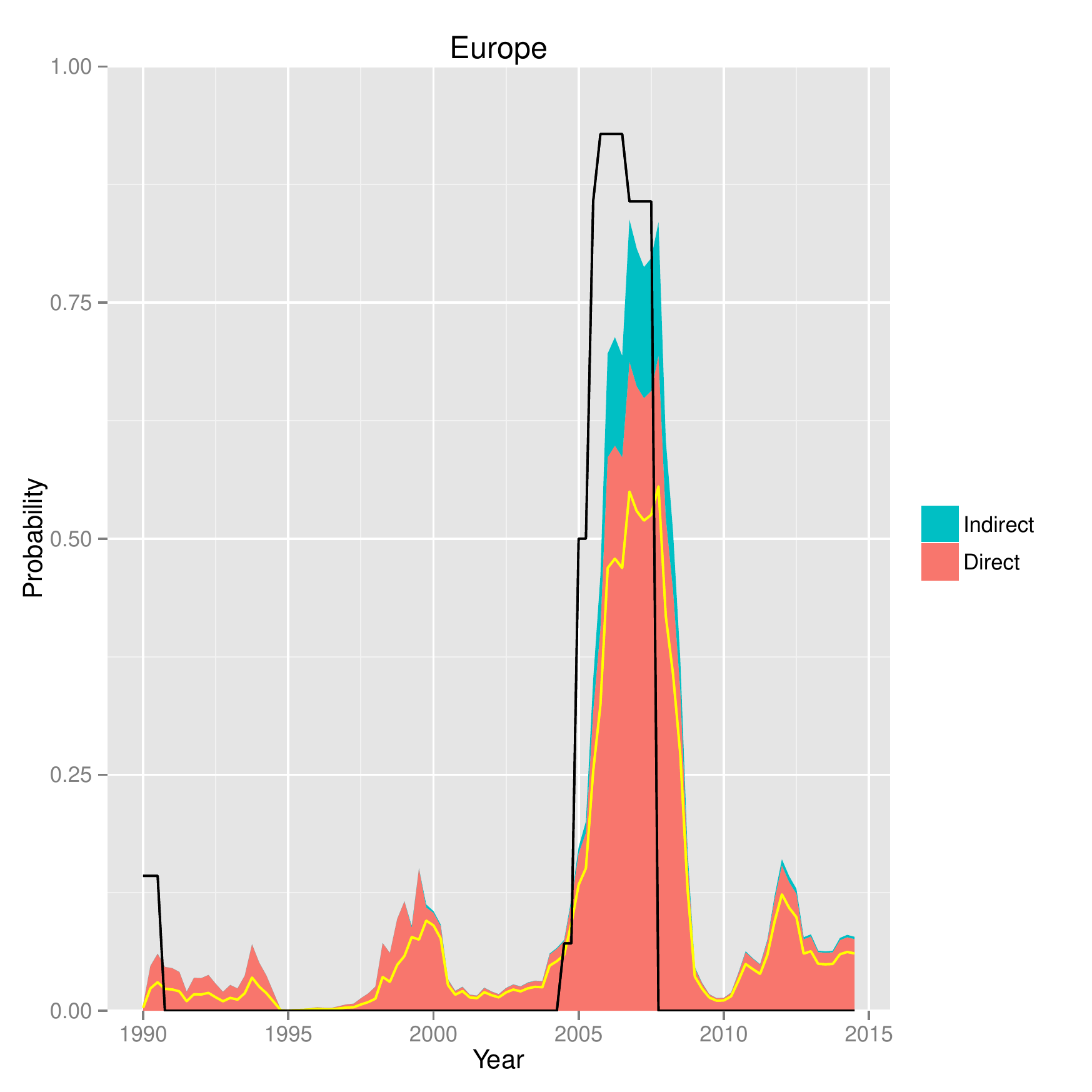}\protect\caption{\label{fig:RiskRank-for-Europe}RiskRank for Europe with the yellow
line representing the weighted average of the individual risk of countries,
while the black line depicts the proportion of countries being in
a pre-crisis state in a given quarter}
\end{figure}

RiskRank also allows to measure vulnerability at a European level
by aggregating the country-specific measures. Figure \ref{fig:RiskRank-for-Europe}
provides an example of model output for Europe with the individual
model and RiskRank. In the figure, the European risk level is decomposed
into direct and indirect impact of countries (red and blue stacks),
whereas the yellow line shows the weighted average of the individual
risk of countries and the black line the proportion of countries being
in a pre-crisis state in a given quarter. Thus, the figure particularly
depicts the increase in RiskRank when connected countries exhibit
larger individual Risk (blue and red stacks vis-á-vis yellow line).
Likewise, the figure also shows that unconditional probabilities (black
line) do exceed a simple weighted average of individual probabilities,
which indeed also points to the need for an aggregation function accounting
for other factors.

\subsection{Robustness of RiskRank}

For a more detailed comparison of predictive performance, Tables \ref{tab:Performance_individual}
and\ref{tab:Performance-aggregated} lists a range of previously introduced
performance measures for the two above models. Further, we test the
robustness of the above exercise when varying model specification.
To check the sensitivity of the aggregation procedure with respect
to the forecast horizon, we perform the analysis with two other scenarios:
(\emph{i}) 5--8 and (\emph{ii}) 5--16 quarters prior to crisis events.
As can be seen in Tables \ref{tab:Country_Results-1} and \ref{tab:Country_Results-2},
RiskRank still results in higher AUC values and relative usefulness
compared to the individual risk model. Naturally, the wider the forecast
horizon is, the worse the predictions are, but the aggregated risk
level used in RiskRank provides a more accurate estimation with clear
improvements for higher values of the preference parameter $\mu$
(above $0.3$).

\begin{table}[H]
\protect\caption{\label{tab:Performance_individual}Performance measures based on individual
probabilities}

\centering{}{\small{}}%
\begin{tabular}{cccccccccc}
{\small{}$\mu$} & {\small{}TP} & {\small{}TN} & {\small{}FP} & {\small{}FN} & {\small{}Precision (C)} & {\small{}Recall (C)} & {\small{}Precision (T)} & {\small{}Recall (T)} & {\small{}Accuracy}\tabularnewline
\hline 
\hline 
{\small{}0.0} & {\small{}0} & {\small{}1146} & {\small{}1} & {\small{}139} & {\small{}0.00} & {\small{}0.00} & {\small{}89.18} & {\small{}99.91} & {\small{}89.11}\tabularnewline
\hline 
{\small{}0.1} & {\small{}0} & {\small{}1146} & {\small{}1} & {\small{}139} & {\small{}0.00} & {\small{}0.00} & {\small{}89.18} & {\small{}99.91} & {\small{}89.11}\tabularnewline
\hline 
{\small{}0.2} & {\small{}0} & {\small{}1146} & {\small{}1} & {\small{}139} & {\small{}0.00} & {\small{}0.00} & {\small{}89.18} & {\small{}99.91} & {\small{}89.11}\tabularnewline
\hline 
{\small{}0.3} & {\small{}30} & {\small{}1138} & {\small{}9} & {\small{}109} & {\small{}76.92} & {\small{}21.58} & {\small{}91.26} & {\small{}99.22} & {\small{}90.82}\tabularnewline
\hline 
{\small{}0.4} & {\small{}30} & {\small{}1138} & {\small{}9} & {\small{}109} & {\small{}76.92} & {\small{}21.58} & {\small{}91.26} & {\small{}99.22} & {\small{}90.82}\tabularnewline
\hline 
{\small{}0.5} & {\small{}30} & {\small{}1138} & {\small{}9} & {\small{}109} & {\small{}76.92} & {\small{}21.58} & {\small{}91.26} & {\small{}99.22} & {\small{}90.82}\tabularnewline
\hline 
{\small{}0.6} & {\small{}98} & {\small{}1052} & {\small{}95} & {\small{}41} & {\small{}50.78} & {\small{}70.50} & {\small{}96.25} & {\small{}91.72} & {\small{}89.42}\tabularnewline
\hline 
{\small{}0.7} & {\small{}113} & {\small{}1028} & {\small{}119} & {\small{}26} & {\small{}48.71} & {\small{}81.29} & {\small{}97.53} & {\small{}89.63} & {\small{}88.72}\tabularnewline
\hline 
{\small{}0.8} & {\small{}116} & {\small{}1018} & {\small{}129} & {\small{}23} & {\small{}47.35} & {\small{}83.45} & {\small{}97.79} & {\small{}88.75} & {\small{}88.18}\tabularnewline
\hline 
{\small{}0.9} & {\small{}121} & {\small{}997} & {\small{}150} & {\small{}18} & {\small{}44.65} & {\small{}87.05} & {\small{}98.23} & {\small{}86.92} & {\small{}86.94}\tabularnewline
\hline 
{\small{}1.0} & {\small{}139} & {\small{}0} & {\small{}1147} & {\small{}0} & {\small{}10.81} & {\small{}1.00} & {\small{}-} & {\small{}0.00} & {\small{}10.81}\tabularnewline
\end{tabular}{\small \par}
\end{table}

\begin{table}[H]
\protect\caption{\label{tab:Performance-aggregated}Performance measures based on aggregated
probabilities}

\centering{}{\small{}}%
\begin{tabular}{cccccccccc}
{\small{}$\mu$} & {\small{}TP} & {\small{}TN} & {\small{}FP} & {\small{}FN} & {\small{}Precision (EW)} & {\small{}Recall (EW)} & {\small{}Precision (T)} & {\small{}Recall (T)} & {\small{}Accuracy}\tabularnewline
\hline 
\hline 
{\small{}0.0} & {\small{}0} & {\small{}1146} & {\small{}1} & {\small{}139} & {\small{}0.00} & {\small{}0.00} & {\small{}89.18} & {\small{}99.91} & {\small{}89.11}\tabularnewline
\hline 
{\small{}0.1} & {\small{}0} & {\small{}1146} & {\small{}1} & {\small{}139} & {\small{}0.00} & {\small{}0.00} & {\small{}89.18} & {\small{}99.91} & {\small{}89.11}\tabularnewline
\hline 
{\small{}0.2} & {\small{}0} & {\small{}1146} & {\small{}1} & {\small{}139} & {\small{}0.00} & {\small{}0.00} & {\small{}89.18} & {\small{}99.91} & {\small{}89.11}\tabularnewline
\hline 
{\small{}0.3} & {\small{}30} & {\small{}1138} & {\small{}9} & {\small{}109} & {\small{}76.92} & {\small{}21.58} & {\small{}91.26} & {\small{}99.22} & {\small{}90.82}\tabularnewline
\hline 
{\small{}0.4} & {\small{}30} & {\small{}1138} & {\small{}9} & {\small{}109} & {\small{}76.92} & {\small{}21.58} & {\small{}91.26} & {\small{}99.22} & {\small{}90.82}\tabularnewline
\hline 
{\small{}0.5} & {\small{}45} & {\small{}1127} & {\small{}20} & {\small{}94} & {\small{}69.23} & {\small{}32.37} & {\small{}92.30} & {\small{}98.96} & {\small{}91.14}\tabularnewline
\hline 
{\small{}0.6} & {\small{}114} & {\small{}1055} & {\small{}92} & {\small{}25} & {\small{}55.34} & {\small{}82.01} & {\small{}97.69} & {\small{}91.98} & {\small{}90.90}\tabularnewline
\hline 
{\small{}0.7} & {\small{}114} & {\small{}1055} & {\small{}92} & {\small{}25} & {\small{}55.34} & {\small{}82.01} & {\small{}97.69} & {\small{}91.98} & {\small{}90.90}\tabularnewline
\hline 
{\small{}0.8} & {\small{}114} & {\small{}1055} & {\small{}92} & {\small{}25} & {\small{}55.34} & {\small{}82.01} & {\small{}97.69} & {\small{}91.98} & {\small{}90.90}\tabularnewline
\hline 
{\small{}0.9} & {\small{}122} & {\small{}998} & {\small{}149} & {\small{}17} & {\small{}45.02} & {\small{}87.77} & {\small{}98.33} & {\small{}87.01} & {\small{}87.09}\tabularnewline
\hline 
{\small{}1.0} & {\small{}139} & {\small{}0} & {\small{}1147} & {\small{}0} & {\small{}10.81} & {\small{}1.00} & {\small{}-} & {\small{}0.00} & {\small{}10.81}\tabularnewline
\end{tabular}{\small \par}
\end{table}

\begin{table}[H]
\protect\caption{\label{tab:Country_Results-1}Signaling performance of individual
models and RiskRank for countries with forecast horizon 5--8 quarters
prior to crisis events.}

\centering{}%
\begin{tabular}{ccccc}
 & \multicolumn{2}{c}{{\small{}Individual}} & \multicolumn{2}{c}{{\small{}RiskRank}}\tabularnewline
{\small{}$\mu$} & \textbf{\small{}$U_{r}(\mu)$} & {\small{}AUC} & \textbf{\small{}$U_{r}(\mu)$} & {\small{}AUC}\tabularnewline
\hline 
\hline 
{\small{}0.0} & {\small{}0 \%} & {\small{}0.927} & {\small{}0 \%} & {\small{}0.935}\tabularnewline
{\small{}0.1} & {\small{}-11 \%} & {\small{}0.927} & {\small{}-11 \%} & {\small{}0.935}\tabularnewline
{\small{}0.2} & {\small{}-5 \%} & {\small{}0.927} & {\small{}-5 \%} & {\small{}0.935}\tabularnewline
{\small{}0.3} & {\small{}-3 \%} & {\small{}0.927} & {\small{}-3 \%} & {\small{}0.935}\tabularnewline
{\small{}0.4} & {\small{}-2 \%} & {\small{}0.927} & {\small{}1 \%} & {\small{}0.935}\tabularnewline
{\small{}0.5} & {\small{}4 \%} & {\small{}0.927} & {\small{}7 \%} & {\small{}0.935}\tabularnewline
{\small{}0.6} & {\small{}11 \%} & {\small{}0.927} & {\small{}16 \%} & {\small{}0.935}\tabularnewline
{\small{}0.7} & {\small{}22 \%} & {\small{}0.927} & {\small{}34 \%} & {\small{}0.935}\tabularnewline
{\small{}0.8} & {\small{}47 \%} & {\small{}0.927} & {\small{}55 \%} & {\small{}0.935}\tabularnewline
{\small{}0.9} & {\small{}70 \%} & {\small{}0.927} & {\small{}76 \%} & {\small{}0.935}\tabularnewline
{\small{}1.0} & {\small{}0 \%} & {\small{}0.927} & {\small{}0 \%} & {\small{}0.935}\tabularnewline
\end{tabular}
\end{table}

\begin{table}[H]
\protect\caption{\label{tab:Country_Results-2}Signaling performance of individual
models and RiskRank for countries with forecast horizon 5--16 quarters
prior to crisis events.}

\centering{}%
\begin{tabular}{ccccc}
 & \multicolumn{2}{c}{{\small{}Individual}} & \multicolumn{2}{c}{{\small{}RiskRank}}\tabularnewline
{\small{}$\mu$} & \textbf{\small{}$U_{r}(\mu)$} & {\small{}AUC} & \textbf{\small{}$U_{r}(\mu)$} & {\small{}AUC}\tabularnewline
\hline 
\hline 
{\small{}0.0} & {\small{}0 \%} & {\small{}0.861} & {\small{}0 \%} & {\small{}0.866}\tabularnewline
{\small{}0.1} & {\small{}-5 \%} & {\small{}0.861} & {\small{}-5 \%} & {\small{}0.866}\tabularnewline
{\small{}0.2} & {\small{}-2 \%} & {\small{}0.861} & {\small{}-2 \%} & {\small{}0.866}\tabularnewline
{\small{}0.3} & {\small{}5 \%} & {\small{}0.861} & {\small{}5 \%} & {\small{}0.866}\tabularnewline
{\small{}0.4} & {\small{}9 \%} & {\small{}0.861} & {\small{}9 \%} & {\small{}0.866}\tabularnewline
{\small{}0.5} & {\small{}11 \%} & {\small{}0.861} & {\small{}15 \%} & {\small{}0.866}\tabularnewline
{\small{}0.6} & {\small{}25 \%} & {\small{}0.861} & {\small{}30 \%} & {\small{}0.866}\tabularnewline
{\small{}0.7} & {\small{}38 \%} & {\small{}0.861} & {\small{}41 \%} & {\small{}0.866}\tabularnewline
{\small{}0.8} & {\small{}50 \%} & {\small{}0.861} & {\small{}50 \%} & {\small{}0.866}\tabularnewline
{\small{}0.9} & {\small{}48 \%} & {\small{}0.861} & {\small{}46 \%} & {\small{}0.866}\tabularnewline
{\small{}1.0} & {\small{}0 \%} & {\small{}0.861} & {\small{}0 \%} & {\small{}0.866}\tabularnewline
\end{tabular}
\end{table}

\section{Conclusion}

Estimating systemic risk is considered nowadays as one of the most
prominent tasks as a consequence of the recent financial crisis. In
this paper we introduced a new approach for assessing systemic risk
motivated by advances in the domains of aggregation operators and
network analysis. In the proposed approach, we combine risk indicators
of entities in the financial markets, consequently modelling cyclical
and cross-sectional systemic risk using our approach. As the result
of an aggregation process, we obtain not only the aggregated risk
levels of individual entities in a system and also the system as a
whole, but the risk is decomposed into individual, direct and indirect
components. While the proposed approach is motivated by the theory
of aggregation operators and network analysis, it is closely related
to, and from some perspectives expands upon the Shapley-index approach
frequently used in economics, specifically in systemic risk analysis. 

The approach is exemplified using the case of estimating systemic
risk in a European setting. In the example, we provide an estimation
of systemic risk from country-level risk and cross-border linkages.
The out-of-sample performance of the approach in this case illustrates
how aggregating risk levels in the network representation can improve
on a traditional estimation method. This paper is to be extended with
also bank-level results, where similar aggregation procedures allow
estimating upward in the hierarchy from bank-level individual risk
and linkages. \newpage{}

\section*{\textmd{\small{}\renewcommand\refname{References}\bibliographystyle{plainnat}
\bibliography{References/references}
}}
\end{document}